\documentclass[aps,pre,twocolumn,superscriptaddress]{revtex4}
\usepackage[usenames,dvipsnames]{color}
\usepackage{amssymb,amsfonts,amsmath}
\usepackage{graphicx}
\usepackage{mathrsfs}
\usepackage{eufrak}
\usepackage{bm}
\usepackage{lmodern}

\usepackage[unicode=true,pdfusetitle, bookmarks=true,bookmarksnumbered=false,bookmarksopen=false, breaklinks=false,pdfborder={0 0 1},backref=false]{hyperref}
\hypersetup{colorlinks=true,
  linkcolor=black,
  anchorcolor=black,
  citecolor=black,
  urlcolor=black
}

\usepackage{breakurl}
\usepackage{notes2bib}

\makeatletter
\@ifundefined{textcolor}{}
{
  \definecolor{ACK}{gray}{0}
  \definecolor{WHITE}{gray}{1}
  \definecolor{RED}{rgb}{1,0,0}
  \definecolor{GREEN}{rgb}{0,1,0}
  \definecolor{BLUE}{rgb}{0,0,1}
  \definecolor{CYAN}{cmyk}{1,0,0,0}
  \definecolor{MAGENTA}{cmyk}{0,1,0,0}
  \definecolor{YELLOW}{cmyk}{0,0,1,0}
}

\usepackage{graphics}
\usepackage{epsfig}
\usepackage{epsf}
\usepackage{esint}
\usepackage{verbatim}
\usepackage{microtype}

\DeclareFontFamily{OT1}{pzc}{}
\DeclareFontShape{OT1}{pzc}{m}{it}{<-> s * [1.40] pzcmi7t}{}
\DeclareMathAlphabet{\mathpzc}{OT1}{pzc}{m}{it}

\DeclareBoldMathCommand\boldlangle{\left\langle}
\DeclareBoldMathCommand\boldrangle{\right\rangle}

\providecommand{\nn}{\nonumber}
\providecommand{\be}{\begin{equation}}
  \providecommand{\ee}{\end{equation}}
\providecommand{\bea}{\begin{eqnarray}}
  \providecommand{\eea}{\end{eqnarray}}
\providecommand{\beas}{\begin{eqnarray*}}
  \providecommand{\eeas}{\end{eqnarray*}}

\providecommand{\beni}{\begin{equation*}}
  \providecommand{\eeni}{\end{equation*}}

\providecommand{\bw}{\begin{widetext}}
  \providecommand{\ew}{\end{widetext}}

\newcommand{\vast}{\bBigg@{2}}
\newcommand{\Vast}{\bBigg@{3}}

\makeatother

\DeclareSymbolFont{mathscrUC}{U}{rsfs}{m}{n}  
\DeclareSymbolFont{mathscrLC}{OT1}{pzc}{m}{n} 

\begin{document}

\title{Entropic forces in a non--equilibrium system:  Flocks of birds}
\author{Michele Castellana}
\affiliation{Joseph Henry Laboratories of Physics and Lewis--Sigler Institute for Integrative Genomics, Princeton University, Princeton, New Jersey 08544, United States}
\author{William Bialek}
\affiliation{Joseph Henry Laboratories of Physics and Lewis--Sigler Institute for Integrative Genomics, Princeton University, Princeton, New Jersey 08544, United States}
\affiliation{Initiative for the Theoretical Sciences, The Graduate Center, City University of New York, 365 Fifth Ave., New York, New York 10016, United States}
\author{Andrea Cavagna}
\affiliation{Initiative for the Theoretical Sciences, The Graduate Center, City University of New York, 365 Fifth Ave., New York, New York 10016, United States}
\affiliation{Istituto dei Sistemi Complessi, Consiglio Nazionale delle Ricerche, Rome, Italy and Dipartimento di Fisica, Universit\`a Sapienza, Rome, Italy}
\author{Irene Giardina}
\affiliation{Initiative for the Theoretical Sciences, The Graduate Center, City University of New York, 365 Fifth Ave., New York, New York 10016, United States}
\affiliation{Istituto dei Sistemi Complessi, Consiglio Nazionale delle Ricerche, Rome, Italy and Dipartimento di Fisica, Universit\`a Sapienza, Rome, Italy}

\begin{abstract}
When birds come together to form a flock, the distribution of their individual velocities narrows around the mean velocity of the flock.  We argue that, in a broad class of models for the joint distribution of positions and velocities, this narrowing generates an entropic force that opposes the cohesion of the flock.  The strength of this force depends strongly on the nature of the interactions among birds: if birds are coupled to a fixed number of neighbors, the entropic forces are weak, while if they couple to all other birds within a fixed distance, the entropic forces are sufficient to tear a flock apart.  Similar entropic forces should occur in other non--equilibrium systems.  For the joint distribution of protein structures and amino--acid sequences, these forces favor the occurrence of ``highly designable'' structures.
\end{abstract}

\maketitle


\section{Introduction}\label{sec4}

Entropic forces are a familiar concept in equilibrium statistical mechanics.  From the ideal gas to the elasticity of random polymers and the effective forces  between molecules in solution, we know that changing the entropy of a system generates a force that is just as ``real'' as the forces that result from changes in energy.  Does this intuition carry over into complex, non--equilibrium systems?

Consider a flock of birds, or a school of fish.  As the animals come close to one another, they interact in ways that cause their velocities to align.  If we imagine constructing the joint distribution of velocities for all the birds in the flock, alignment means that the entropy of this distribution goes down.  Is there a resulting entropic force that pushes the birds apart, allowing the entropy to increase?  If this were an equilibrium system, the answer would be yes.  But this is not an equilibrium system, by any means.  Are there still entropic forces?

In recent years there has been renewed interest in the use of maximum entropy methods to describe the collective behavior of biological networks, with applications spanning scales from the network of amino acids in a family of proteins \cite{bialek2007rediscovering,seno2008,weigt2009identification,marks2011protein,sulkowska+al2012}, to biochemical and genetic networks \cite{lezon2006using,tkacik2007}, networks of neurons  \cite{schneidman2006weak,tkacik2006ising,shlens2006,tang2008,tkacik2009spin,Ohiorhenuan2010,ganmor2011,sdme2013,tkacik2014critical}, and flocks of birds \cite{bialek2012statistical,bialek2014speed}.  The idea of the maximum entropy method is to construct the least structured model of a system that is consistent with certain measured average properties \cite{jaynes57}.  If the only quantity that we measure is the energy, then constructing the maximum entropy distribution is exactly the construction of the thermal equilibrium, Boltzmann distribution.  But in these complex biological systems, the quantities that we can measure are not the energy, and typically there are many such quantities.  The maximum entropy distribution that we construct then is not at all an equilibrium distribution for the system we are studying, although it is mathematically equivalent to the equilibrium statistical mechanics of some other system.  

In the maximum entropy framework, it is relatively easy to show that even non--equilibrium systems are subject to entropic forces.  In the context of flocking, this really does mean that birds are repelled by the loss of entropy associated with their mutual orientation.  But, in detail, we will see that this effect depends dramatically on the nature of the interactions and ordering in the system.  If we imagine that the flight direction of individual birds maintains a certain level of correlation with the average direction of its $n_c$ nearest neighbors (``topological interactions'' \cite{ballerini2008interaction}), then the entropic forces are weak, and (in a sense that we will make precise) flocks can cohere even without explicit forces holding them together.  On the other hand, if flocks are characterized by correlation between a bird and  its neighbors within some characteristic distance $r_c$ (``metric interactions''), then the entropic effects are strong enough that almost all reasonable flocks will be broken into multiple disconnected pieces unless there are other explicitly cohesive forces.  Although such forces surely exist, observations on real flocks of starlings show that the positional correlations among birds are weak \cite{cavagna2008gofr}, so if there are strong repulsive forces from the entropy of flight directions, these would have to be finely balanced by attractive interactions.  Such fine tuning is unnecessary in the case of topological interactions.

Although we can make analytic progress on evaluating the entropic forces that result from directional ordering in the flock, computing the impact of these forces on the distribution of the birds' positions must be done numerically.  This becomes challenging for large flocks, and so we have formulated a simpler problem in which the birds live not in the full three dimensional space, but on a a graph, and we carry out Monte Carlo (MC) simulations in the space of these graphs.  We can see that the two problems have similar structure, and in particular the differences between metric and topological interactions arise in both cases; the graph model allows us to follow these difference out to larger systems.

\section{Entropic forces in maximum entropy models}\label{entforces}

The essential intuition that we use in building maximum entropy models for  flocks of birds is that the dominant interactions are local, and hence if we want to characterize the nature of order in the flock we should measure the degree of correlation between the flight velocities $\vec{v}_i$ of birds and their near neighbors \cite{bialek2012statistical,bialek2014speed}.  To be concrete, we will consider only normalized velocities neglecting variations in speed, so each bird $i$ is described by a unit vector $\vec{s}_i \equiv \vec{v}_i / \left| \vec{v}_i \right|$ and is located at position $\vec{x}_i$. Around  bird $i$ we define a neighborhood ${\cal N}_i$, and within this neighborhood there are $n_i$ neighbors---details of how this neighborhood is defined are discussed below.  To measure the correlation of each bird's direction with the average of its neighbors,  we compute
\begin{equation}
C_{\rm int} = {1\over N} \sum_{i=1}^N {\bigg\langle} \vec{s}_i  \cdot \vast( {1\over {n_i}}\sum_{j\in {\cal N}_i} \vec{s}_j \vast){\bigg\rangle},
\end{equation}
where the sum over index $i$ runs over birds with $n_i \neq 0$, and $\langle{\phantom{i}}\rangle$ denotes the average with respect to the joint distribution of  directions $s \equiv \{ \vec{s}_i \}$. 
It will be crucial in what follows that, although $C_{\rm int}$ depends explicitly on flight directions, it depends implicitly on the birds' positions $x \equiv \{\vec{x}_i\}$.
We can make this explicit by defining an adjacency matrix $n_{ij}(x)$ such that $n_{ij} = 1$ if $j\in{\cal N}_i$, and zero otherwise.    Then we have
\begin{eqnarray}
C_{\rm int} &=& {1\over N} \sum_{i=1}^N \sum_{j=1}^N{\bigg\langle} {{n_{ij}(x)}\over{n_i(x)}} \vec{s}_i  \cdot\vec{s}_j {\bigg\rangle} ,
\label{C2}\\
n_i(x) &=& \sum_{j=1}^N n_{ij}(x) .
\end{eqnarray}

If the local correlations $C_{\rm int}$ characterize the nature of ordering in the flock, then we should obtain a good approximation to the full, joint distribution of flight directions by building the maximum entropy distribution consistent with the value  of $C_{\rm int}$ observed in real flocks.  Indeed, this works:  the maximum entropy distribution that matches $C_{\rm int}$ provides accurate, parameter--free predictions for the behavior of two-- and four--point correlations as a function of distance, out to length scales comparable to the size of the flock itself \cite{bialek2012statistical}.  

There are two different points of view that we can take on the maximum entropy construction.  In the first view, the positions of the birds are known, and we are constructing the distribution of flight directions given these positions.  This maximum entropy distribution is
\begin{equation}
P(s|x) = {1\over {Z(x; J)}} \exp\Vast[ J \sum_{i=1}^N \sum_{j=1}^N {{n_{ij}(x)}\over{n_i(x)}} \vec{s}_i  \cdot\vec{s}_j\Vast] ,\label{Psgivenx}
\vspace{1mm}
\end{equation}
where, as usual, the partition function is given by
\begin{equation}
Z(x; J) = \int ds \exp\left[ J \sum_{i=1}^N \sum_{j=1}^N {{n_{ij}(x)}\over{n_i(x)}} \vec{s}_i  \cdot\vec{s}_j\right] ,
\label{Z1}
\end{equation}
and $\int ds \equiv \int d\vec{s}_1 \cdots d\vec{s}_N$. 
The parameter $J$ is determined by the condition that $C_{\rm int}$ computed from this distribution matches what we observe for the real flock, $C_{\rm int}^{\rm obs}$ \cite{bialek2012statistical}, and this is equivalent to solving the equation
\begin{equation}
{{\partial \ln Z(x; J)}\over{\partial J}} =  N C_{\rm int}^{\rm obs}.
\end{equation}
This description of flight directions given positions is useful in part because the neighbor relations among birds in the flock change slowly compared to flight directions \cite{cavagna2013diffusion}, and fluctuations in $C_{\rm int}$ from moment to moment in a single flocking event are small.

In the second view, we imagine that we observe a flock for a very long time, long enough for the birds  to rearrange substantially, exchanging neighbors.  Then when we compute the average involved in defining $C_{\rm int}$, Eq. (\ref{C2}), we are averaging not just over flight directions but also over positions.  Now we can ask for the maximum entropy distribution of (jointly) positions and flight directions that is consistent with the $C_{\rm int}^{\rm obs}$, and the answer is
\begin{equation}
P(x, s) = {1\over{Z_0(J)}} \exp\left[ J \sum_{i=1}^N \sum_{j=1}^N {{n_{ij}(x)}\over{n_i(x)}} \vec{s}_i  \cdot\vec{s}_j\right] .
\label{pxs1}
\end{equation}
Of course we might know more about the flock than just $C_{\rm int}^{\rm obs}$.  For example, we might have some information about the distribution of pairwise distances between birds, in which case the maximum entropy distribution becomes
\begin{widetext}
\begin{equation}
P(x, s) = {1\over{Z_1(J)}} \exp\left[ -\sum_{i=1}^N \sum_{j=1}^N V(|\vec{x}_i - \vec{x}_j|) +  J \sum_{i=1}^N \sum_{j=1}^N {{n_{ij}(x)}\over{n_i(x)}} \vec{s}_i  \cdot\vec{s}_j\right] ,
\label{pxs2}
\end{equation}
\end{widetext}
where the effective potential $V(r)$ must be tuned to match the distribution of pairwise distances.    

Once we have a model for the joint distribution of positions and velocities, we can integrate out the velocities to give the distribution of positions alone.  We will refer to this as the ``motional distribution'',  $P_{\rm mot}(x)$, since if we start in the simplest case of Eq. (\ref{pxs1}) all the nontrivial structure of this distribution arises from the motion of the birds.  We have
\begin{equation}
P_{\rm mot}(x) \equiv \int ds P(x, s) = {{Z(x; J)}\over{Z_0(J)}} .
\label{pmot1}
\end{equation}
Thus, allowing ourselves the usual language of statistical mechanics, the free energy $F(x) = -\ln Z(x;J)$ acts as an effective potential for the flock, 
\begin{equation}
P_{\rm mot}(x) \propto e^{-F(x)}.
\end{equation}
Notice that if the flock is perfectly ordered, so that all $\vec{s}_i$ are equal, then the exponential in Eq. (\ref{pxs1}) is just $JN$, independent of $x$.  In fact, real flocks are highly polarized, and we can compute $Z(x; J)$ with an expansion around this perfectly ordered state \cite{bialek2012statistical,bialek2014speed} by means of the spin--wave approximation---a method used in solid-state physics to study perturbations in fully ordered ferromagnetic states \cite{dyson1956general}.  In this approximation, the free energy $F(x)$ is dominated by the entropy of the fluctuations in the flight directions, so that gradients in this free energy constitute an entropic force on the birds' spatial configuration.  If we have other constraints on the distribution as in Eq. (\ref{pxs2}), then the free energy $F(x)$ of the flight directions just adds another term to the effective potential, as usual.

We conclude this discussion with a cautionary remark about the interpretation of maximum entropy models, and their relation to equilibrium statistical physics.  When we look at Eq. (\ref{pxs2}), it is tempting to note the equivalence with a Boltzmann distribution and interpret the term in the exponential as the Hamiltonian of the system, and we will sometimes lapse into this language ourselves.  But this mapping obviously does not mean that the effective Hamiltonian is really the energy of the system, nor does it even mean that the dynamics correspond to Brownian motion in the effective potential.  In more biologically motivated models of flocks, one speaks of ``social forces'' that drive cohesion and orientational ordering \cite{aoki82,viscek+al_95,couzin+al02,gregoire+al03,hildebrandt2010}, and one might tempted to identify these social forces with derivatives of the two terms in the Hamiltonian, but this need not be correct.  As is well known, there are infinitely many dynamical processes that can give rise to the same stationary distribution.  The maximum entropy method aims to characterize this distribution directly, incorporating only the structure needed to match a small set of experimental observations.  The choice of which observations to match is based on intuition, and must be tested by checking that the resulting maximum entropy distributions actually provide an accurate description of the system, as in Ref.\cite{bialek2012statistical}.  The equivalence to equilibrium statistical--physics models means that we can carry over much of our what we know about expectation values, correlation functions and, as we have seen, entropic forces.  But we cannot jump from this probabilistic description back to a model of the underlying dynamics.\\

\section{The motional free energy}

Our task now is to compute the partition function in Eq. (\ref{Z1}).  It is useful to note that this can be rewritten more symmetrically, as
\begin{eqnarray}
Z(x; J) &=& \int ds \exp\Bigg[  \sum_{i,j=1}^N J_{ij}(x) \, \vec{s}_i \cdot \vec{s}_j \Bigg],
\label{eq4}\\
\label{eq55}
J_{ij}(x) &=& \frac{J}{2} \left[ \frac{n_{ij}(x)}{n_i(x)} +  \frac{n_{ji}(x)}{n_j(x)} \right].
\end{eqnarray} 
A configuration $x$ defines a graph $G(x)$ with $N$ vertices, where each nonzero element of  the adjacency matrix $n_{ij}(x)$ corresponds to an edge between vertices $i$ and $j$ in $G$. We denote by $k$ the number of connected components in  $G$,
and by $N_l$ the number of birds in the $l$-th connected component, with $l=1, \ldots, k$. We can relabel the birds so that the matrix $J_{ij}$ consists of $k$ uncoupled blocks.

To evaluate $Z(x;J)$, we are going to use the spin--wave approximation, which is valid at large $J$, such that each connected component of the flock is strongly polarized.  For each block we define the net polarization
\be
\vec{S}_l \equiv \frac{1}{N_l} \sum_{i \in N_l} \vec{s}_i \equiv S_l \hat{n}_l,
\ee
where $S_l$, $\hat{n}_l$ are the norm and the direction of $\vec{S}_l$, respectively, and the sum for $i \in N_l$ runs over all birds in the $l$-th connected component. We now decompose the velocity $\vec{s}_i$, with $i\in N_l$, into components parallel and perpendicular to $\vec{S}_l$
\be \label{eq5}
\vec{s}_i = s^{L}_i \hat{n}_l + \vec{\pi}_i. 
\ee
Substituting into  Eq. (\ref{eq4}), we obtain
\bw
\be \label{eq6}
Z( x;J)  =  \prod_{l=1}^{k} \int d \vec{S}_l \, d^{N_l}s^L \, d^{N_l} \vec{\pi} \prod_{i\in N_l} \frac{\delta\big(s^L_i -\sqrt{1-|\vec{\pi}_i|^2}\big)}{2 \sqrt{1-|\vec{\pi}_i|^2}} \exp\Bigg[ \sum_{i,j\in N_l}^N J_{ij}(x) (s^L_i s^L_j + \vec{\pi}_i \cdot \vec{\pi}_j)  \Bigg] \delta \Bigg(  \vec{S}_l - \frac{1}{N_l} \sum_{i \in N_l} \vec{s}_i \Bigg),
\ee
\ew
where in Eq. (\ref{eq6}) the first Dirac delta results from integration over the unit sphere, we used the fact that $J_{ij}$ is a block matrix, we  inserted a factor of unity
\[
\int d\vec{S}_l \, \delta \Bigg(  \vec{S}_l - \frac{1}{N_l} \sum_{i \in N_l} \vec{s}_i \Bigg) = 1,
\]
and we  rewrote the dot product $\vec{s}_i \cdot \vec{s}_j$ in terms of the $s^L ,\vec{\pi}$ coordinates by using  Eq. (\ref{eq5}).

Guided by the experimental observation that birds in a block fly in directions mostly parallel to $\vec{S}_l$ \cite{bialek2012statistical}, we  assume that the perpendicular velocity components $\vec{\pi}_i$ are small, and we will thus expand the right--hand side of Eq. (\ref{eq6}) in powers of $\vec{\pi}_i$ with the spin--wave approximation. Specifically, we manipulate Eq. (\ref{eq6}) as follows: We neglect
$|\vec{\pi}_i|^2$ in the square root in the denominator, we integrate with respect to $\{ s^L_i \}$, and we expand the term in parentheses in the exponent to leading order as
\[
s^L_i s^L_j = \sqrt{1-\left|\vec{\pi}_i\right|^2} \sqrt{1-\left|\vec{\pi}_j\right|^2} \approx 1 - \frac{|\vec{\pi}_i|^2}{2} - \frac{|\vec{\pi}_j|^2}{2}. 
\]
In addition, we rewrite the Dirac delta function in the last line of Eq. (\ref{eq6})  in terms of the $s^L , \vec{\pi}$ coordinates as
\bw
\be \label{eq8}
\delta \Bigg(  \vec{S}_l - \frac{1}{N_l} \sum_{i \in N_l} \vec{s}_i \Bigg) = 
\delta\Bigg(S_l - \frac{1}{N_l} \sum_{i \in N_l} s^L_i\Bigg)   \delta \Bigg( \frac{1}{N_l} \sum_{i \in N_l} \vec{\pi}_i \Bigg), 
\ee
and we perform the integration with respect to $\vec{S}_l$ in spherical coordinates.  The result is 
\be \label{eq9}
Z( x;J)   = \prod_{l=1}^l 4 \pi \int d^{N_l} \vec{\pi} \exp \Bigg[  \sum_{ij \in N_l}  J_{ij}(x) \left(1 - \frac{|\vec{\pi}_i|^2}{2} - \frac{|\vec{\pi}_j|^2}{2} + \vec{\pi}_i \cdot \vec{\pi}_j \right) \Bigg] \delta  \left( \frac{1}{N_l} \sum_{i \in N_l} \vec{\pi}_i \right). 
\ee
Note that the factors of $4 \pi$ in Eq. (\ref{eq9}) arise from the angular integration over all possible directions of the mean velocity $\vec{S}_l$ of the $l$-th connected component: these are explicitly entropic terms.

In what follows, we will drop  the first addend in the exponential in Eq. (\ref{eq9}), because this term  gives rise to a multiplication factor $e^{NJ}$ which is independent of the positional configuration $x$ as noted above.
Then, since  $\vec{\pi}_i $ is a two-dimensional vector, the integral with respect to  $\vec{\pi}_i $ can be rewritten as a product of two identical integrals
\be\label{eq22}
Z(x;J) = \prod_{l=1}^l 4 \pi \Bigg\{ \int d^{N_l} \pi \exp \Bigg[- \sum_{ij \in N_l}  \pi_i \Lambda_{ij}(x) \pi_j \Bigg] \delta  \Bigg( \frac{1}{N_l} \sum_{i \in N_l} {\pi}_i \Bigg) \Bigg\}^2, 
\ee
where $\pi_i$ denotes one component of $\vec{\pi}_i$, and we introduced the Laplacian
\be  \label{eq11}
\Lambda_{ij} (x) \equiv \delta_{ij} \sum_{l =1 }^N J_{li}(x) -  J_{ij}(x);
\ee
since $J_{ij}$ has a block structure, so does $\Lambda_{ij}$.  We denote the eigenvalues and eigenvectors of the $l$-th block of the Laplacian  
by $\{ \lambda^l_p \}$ and $\{ \vec{v}^l_p \}$, respectively, so that
\be
\Lambda_{ij} = \sum_{p=1}^{N_l} v^l_{p, i}  \lambda^l_p   v^l_{p, j}, \;\; i,j \in N_l .
\ee
Summing both sides of Eq. (\ref{eq11}) with respect to $j \in N_l$, we find  that the vector $\vec{v}^l_1 = 1/\sqrt{N_l} (1, \cdots, 1)$ is an eigenvector with eigenvalue $\lambda^l_1 = 0$ \cite{caughman2006kernels}. 
Setting $u_{l,p} \equiv \sum_{i=1}^{N_l} v^l_{p,i}\pi_i$,  we finally obtain 
\be \label{eq26}
Z(x;J)  = 
\prod_{l=1}^l 4 \pi \Bigg\{ \int d^{N_l} u \exp \Bigg[ - \sum_{p = 2}^{N_l}  \lambda^l_p(x)  {u}_{l,p}^2 \Bigg]  \delta  \left( \frac{u_{l,1}}{\sqrt{N_l}} \right) \Bigg\}^2 
=   \prod_{l=1}^l 4 \pi N_l \prod_{p=2}^{N_l} \frac{\pi}{\lambda^l_p(x)}.
\ee
\ew
In what follows, we will consider the impact of the motional free energy alone, uncompensated by any other knowledge of the distribution of the birds' positions.  That is, we will follow Eq. (\ref{pmot1}) and write
\be \label{eq36}
P_{\rm{mot}}(x)  \propto Z(x;J) =    \prod_{l=1}^k 4 \pi N_l  \prod_{p=2}^ {N_l} \frac{\pi}{\lambda^l_p(x)}.
\ee

Equation (\ref{eq36}) tells us that if the flock with $N$ birds consists of one connected cluster, then there are $N-1$ powers of $J/\pi$ in the denominator of $P_{\rm mot}$, since all $\lambda_p^l \propto J$, and an overall factor of $4\pi N$.  If the flock is cut in two halves, then we lose one power of $J/\pi$, and the factor $4\pi N \rightarrow (4\pi N /2)^2$; the net result is to multiply $P_{\rm mot}$ by $ N J$.  This factor is essentially the increase in entropy associated with the creation of a new zero mode in the joint distribution; evidently for large $N$ and large $J$, it favors breaking the flock in half.  There is some subtlety, however, since when we break the flock in half we also shift the spectrum of all the modes in each half, and it is not clear how this balances against the zero mode.  We will see that the answer depends on the nature of the interactions between birds.

\section{Metric vs topological interactions}\label{results2}

In what follows we will consider a flock of $N$ birds in a volume $V$, hence at mean density $\rho = N/V$.    We can imagine the interaction between birds having two very different forms \cite{ballerini2008interaction}.  In the first model, birds interact with  other birds within a characteristic distance $r_c$.  In the second model, birds interact with their $n_c$ nearest neighbors, independent of distance.  The first model is referred to as a ``metric'' interaction, while the second is summarized as a ``topological'' interaction.    We can compare the two models by equating the mean number of interacting neighbors in the metric case with the fixed number of neighbors in the topological case, $n_c = 4\pi \rho \, r_c^3 / 3$.  In both models this is the only relevant dimensionless parameter at large $N$ and $V$.

For a topological network, the adjacency matrix is  
\be \label{eq19}
n_{ij}(x) \equiv \left\{ 
  \begin{array}{ll}
    1 & \textrm{If } j \neq i \textrm{ is amongst the first } n_c\\
    &  \textrm{nearest neighbors of } i \\
    0 & \textrm{Otherwise}
  \end{array}\right. .
\ee
Here we have $n_i = n_c$ for all $i$, and Eq. (\ref{eq55}) takes the simple form 
\be \label{eq13}
J_{ij} (x) = \frac{J}{2 n_c} ( n_{ij}(x) + n_{ji}(x) ).
\ee
For a metric network, the adjacency matrix is
\be \label{eq20}
n_{ij}(x) \equiv \left\{ 
  \begin{array}{ll}
    1 & \textrm{If } j \neq i \textrm{ and }|\vec{x}_j-\vec{x}_i|\leq r_c \\
    0 & \textrm{Otherwise}
  \end{array}\right. ,
\ee
and the interaction matrix is given by Eq. (\ref{eq55}).

In the first models for collective behavior, the metric structure of interactions seemed obvious.  Later on, quantitative studies on flocks of starlings \cite{ballerini2008interaction} led to the idea of topological interactions, and this has been supported by further analyses of these data \cite{bialek2012statistical}. The spatial arrangements of starling flocks are relatively uniform, however, so that the evidence for topological vs. metric interactions is based on the comparison of different flocking events, in which birds assemble at significantly different densities; across many such events, the data are described very accurately by a fixed number of neighbors $n_c$, rather than by a fixed interaction range $r_c$.  In what follows, we will allow model flocks to explore a much wider range of configurations driven by the entropic forces, and so we expect the distinction between topological and metric interactions to become more clear.

\begin{figure*}[tb]
  \centerline{\includegraphics[width = \linewidth]{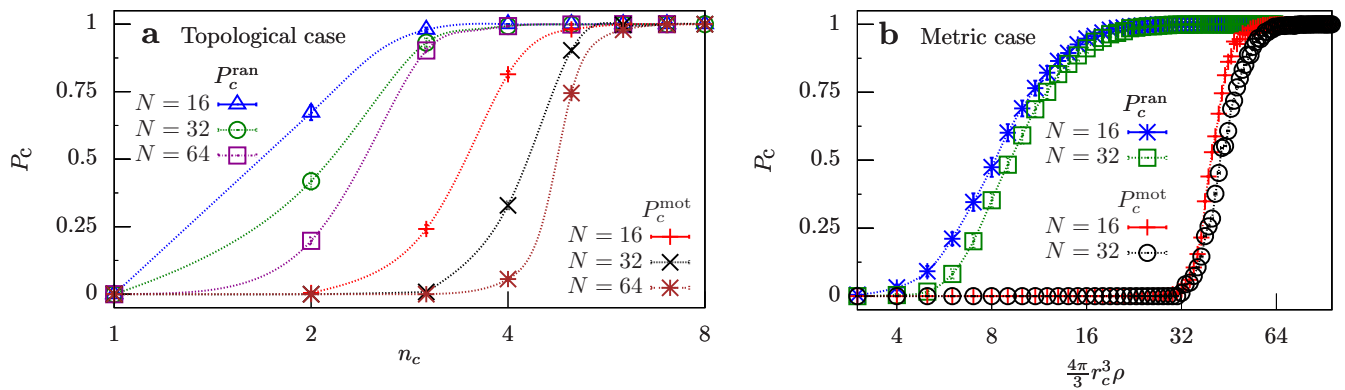}}
  \caption{Probability of finding a connected flock.  Connection probabilities $P_c^{\rm{mot}}$, $P_c^{\rm{ran}}$ from the motional potential and from random positions, respectively, as functions of the number of neighbors $n_c$, both in the topological and metric case. In the metric case, the average number of neighbors $n_c$ is obtained from the interaction range $r_c$ from the relation $n_c = \frac{4\pi} {3} r_c^3\rho$, where $\rho$ is the average density of the flock. Error bars have been estimated with the bootstrap method \cite{newman1999monte}.
    \label{fig5}}
\end{figure*}

\section{Monte Carlo sampling of the motional distribution}\label{results4}

We are interested in exploring the entropic forces generated by the alignment of the birds, as described by Eq. (\ref{eq36}).  Since we can write the distribution of positions analytically, it is natural to generate samples from this distribution with MC simulations.  We will do this directly for both metric and topological interactions in Section \ref{results4.3}.  Although there are clear results, we find that it is difficult to push these simulations to very large $N$, essentially because calculating the eigenvalues of the Laplacian requires $O(N^3)$ operations.  To make progress, in Section \ref{results4.3} we introduce a simpler version of the problem, in which we sample interaction graphs rather than the underlying positions.  While not quite the same problem, we will see consistent results that extend to much larger $N$. The essential point to emerge from both analyses is that topological interactions lead to flocks that stay connected for realistic values of the number of neighbors, while in metric networks the repulsive entropic force is strong enough to rip the flock apart into multiple components for all sensible values of the metric--interaction range.

\subsection{Monte Carlo on positions}\label{results4.1}

Equation (\ref{eq36}) gives us a model for the distribution of birds' positions in a flock with only two parameters:  the strength of the interactions $J$ and their range $n_c$.  Before we study the effects of these interactions on the spatial configurations of the flock, we should start by asking what happens if the $N$ birds are simply in random positions, drawn uniformly throughout a box of volume $V$.  Even in this simple case there is a question about whether the resulting network of interactions---metric or topological---supports a single, connected cluster of birds, thus allowing for the possibility of coherent flocking behavior.

Concretely, for any spatial configuration we can construct the adjacency matrix $n_{ij}(x)$, see Eq. (\ref{eq19}) or (\ref{eq20}), the resulting interaction matrix $J_{ij} (x)$ in Eq. (\ref{eq55}), and finally the Laplacian $\Lambda_{ij} (x)$ in Eq. (\ref{eq11}).  Then we   count the zero modes of the Laplacian:  if there is just one, corresponding to freedom in the overall flight direction, then the flock is connected; if there is more than one, then the flock has broken into disconnected pieces.  

In Fig. \ref{fig5} we show results on the probability $P_c$ of finding a single connected cluster for random configurations, as a function of $n_c$ for metric and topological interactions.  We see that, for topological interactions, connectedness is guaranteed by very modest values of $n_c$.  Perhaps surprisingly, this is not the case for metric interactions.  When interactions are limited to a fixed distance,    even random fluctuations are enough to prevent the formation of a single connected cluster, unless the range of interactions is very large.   This is a hint that metric and topological interactions really are very different.

To see the effects of entropic forces, we need to draw samples from Eq. (\ref{eq36}).  We will view $n_c$ as a parameter to be varied, but we would like to set the strength of interactions to some reasonable value.  The maximum entropy models that we are discussing have been built for real flocks, matching the observed values of $C_{\rm int}$ \cite{bialek2012statistical}, and in our notation this matching determined $J/n_c \sim 16$ \bibnote[note1]{In Eq. (\ref{Psgivenx}), with topological interactions we have $n_i(x) = n_c$ for all $i$, and hence 
\begin{equation}
P(s|x) = {1\over{Z(x;J)}} \exp\left[ {J\over{n_c}}\sum_{i=1}^N \sum_{j=1}^N n_{ij}(x) {\vec s}_i \cdot {\vec s}_j \right] .
\end{equation}
Comparing with Ref. \cite{bialek2012statistical}, we see that what was $J$ in the previous work is $J/n_c$ in the current formulation. In the metric case, we compare Eq. (\ref{Psgivenx})  with the flock configurations in Ref. \cite{bialek2012statistical} with a nearly homogeneous density $\rho$ by setting  $n_i(x) \approx n_c = 4\pi \rho \, r_c^3 / 3$, and we obtain the same relation between the interaction strengths.}

\begin{figure*}[tb]
  \centerline{\includegraphics[width=0.75\linewidth]{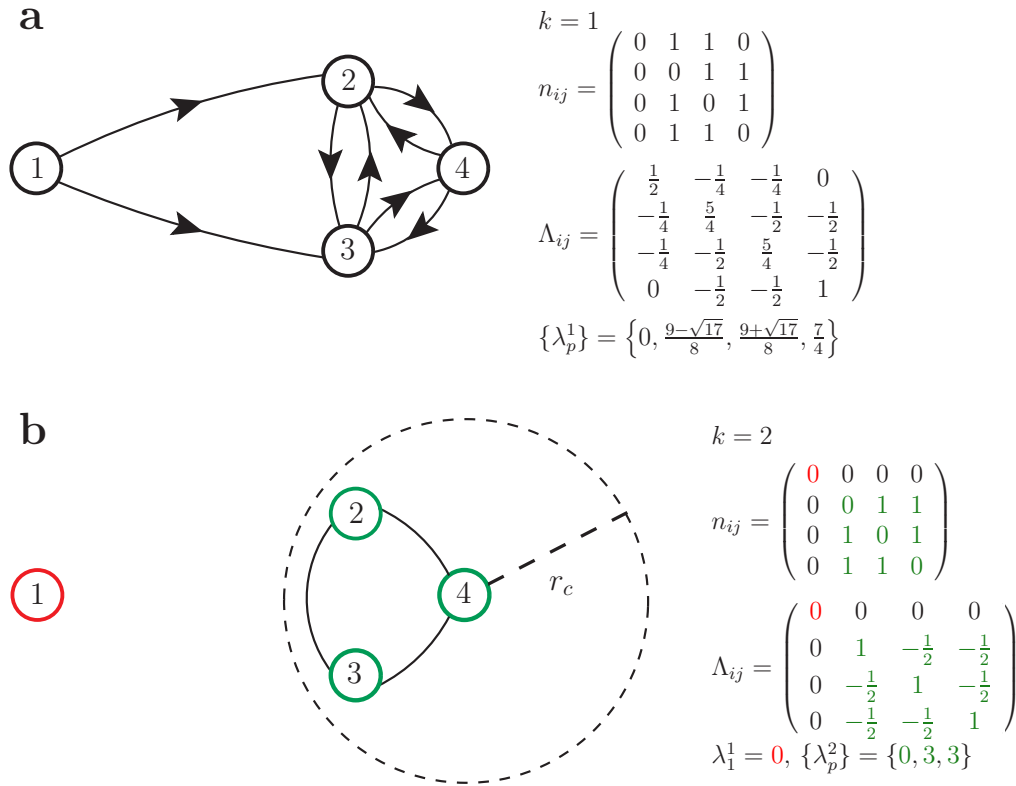}}
  \caption{
    Graphs corresponding to a a two-dimensional  flock with $N =4$ birds and $J=1$. The number of connected components, the adjacency and Laplacian matrices, and the Laplacian eigenvalues are also shown.
    (a) Topological case: Here $n_c = 2$, and the graph $G$ is a directed graph with one connected component. The Laplacian has only one null eigenvalue, corresponding to one connected component in the graph \cite{caughman2006kernels}. 
    (b) Metric case: $r_c$ is the radius of the dashed circle centered around bird $4$, and the graph $G$ is an undirected graph with two connected components.  The first connected component is given by bird $1$ (in red), and the second connected component is given by birds $2,3,4$ (in green). The adjacency and Laplacian matrices are block matrices composed of two blocks corresponding to the first and second connected component depicted in red and green, respectively. The Laplacian has two null eigenvalues, corresponding to two connected components in $G$.  
    \label{fig1}}
\end{figure*}

As we draw samples from $P_{\rm mot}(x)$, we focus on the probability $P_c$ that these spatial configurations correspond to a single connected cluster; results are shown in Fig. \ref{fig5}.  We see that the entropic forces indeed are repulsive, so that interacting flocks are less likely to be connected than random flocks, and this is true at all the values of $N$ and $n_c$ that we can access, for both metric and topological interactions.  But for topological interactions, once $n_c \gtrsim 7$ we find that $P_c$ is essentially equal to one.  In contrast, for the metric interactions there is a very gradual dependence of $P_c$ on $n_c$ that sets in only at large $n_c$.  This effect is dramatic because---in the range of $N$ we can study---the entropic forces are ripping the flock apart even when the range of metric interactions is so large that $n_c > N$.

\subsection{Monte Carlo on graphs}\label{results4.3}

The adjacency matrix $n_{ij}(x)$ defines a graph $G(x)$, as shown for a (very) small flock in Fig. \ref{fig1}.  In the topological case, any positional configuration $x$ corresponds to a graph $G(x)$ with a fixed number $n_c$ of outgoing edges per vertex and thus a fixed total number of edges. Conversely, in the metric case the number of edges connected to a vertex may vary depending on the positional configuration $x$, but the total number of edges $M$ is almost fixed, with   
\be\label{eq18}
M = \sum_{i < j=1}^Nn_{ij}  = \frac{N}{2}  \left(\frac{1}{N} \sum_{i=1}^N n_{i} \right) \approx \frac{N}{2} n_c .
\ee
In the last step we replaced $(1/N )\sum_{i=1}^N n_i \rightarrow n_c$, valid at large $N$ if there are no long--range correlations among the fluctuations in $n_i$.   If we could neglect fluctuations in $n_i$ all together, then the Laplacian matrix $\Lambda_{ij}$ would be proportional to a sparse matrix with integer coefficients, allowing for a much more efficient computation.

These observations suggests a natural way to simplify MC simulations. Instead of sampling the positional configurations $x$, in the topological case we sample all directed graphs with $n_c$ outgoing edges from each vertex, while in the metric case we sample all undirected graphs with $Nn_c/2$ total edges,  assigning to a graph $G$ the probability 
\be \label{por}
\mathpzc{p}_{\rm{mot}}(G)  =   C \prod_{l=1}^k 4 \pi N_l  \prod_{p=2}^ {N_l} \frac{\pi}{\lambda^p_l(G)} ,
\ee
where the $\lambda^p_l$ are calculated in the approximation that $n_i \rightarrow n_c$ in Eq. (\ref{eq55}).  Since the total number of edges is constant both in the topological and metric case, a MC move is now  an edge insertion/deletion, and the impact of these moves on the eigenvalues of the Laplacian can be computed in $O(N^2)$ operations by using the LDL matrix factorization method \cite{davis1999modifying,davis2013user}  (see Appendix \ref{app1}).   As a result, we will be able to study values of $N$ comparable to those of natural flocks, i.e. $N \sim 1000$.

\begin{figure*}[tb]
  \centerline{\includegraphics[width = 0.8\linewidth]{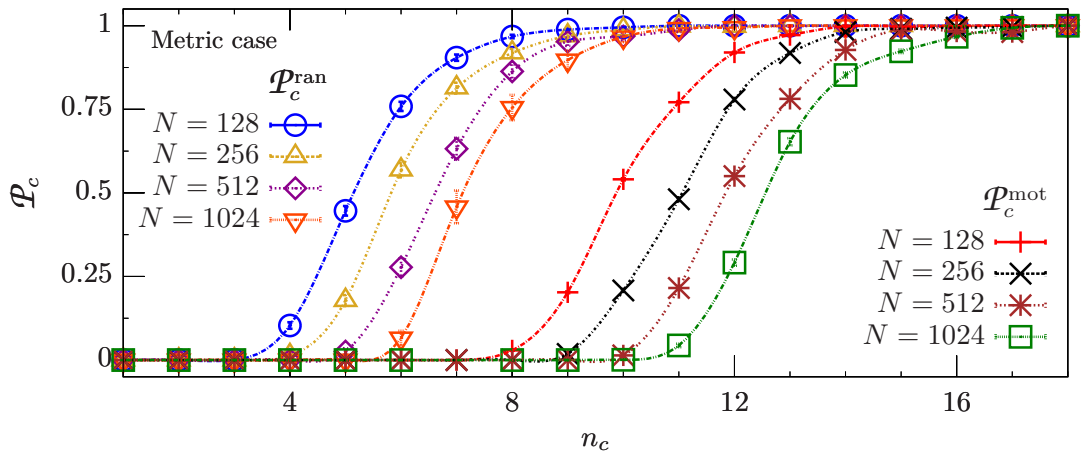}}
  \caption{Probability of finding a connected flock with the simplified Monte Carlo method on graphs: connection probabilities  $\mathpzc{P}_c^{\rm{mot}}$, $\mathpzc{P}_c^{\rm{ran}}$ from the motional potential and from random graphs respectively as functions of the number of neighbors $n_c$ in the metric case. In the topological case, $\mathpzc{P}_c^{\rm{mot}}$ is equal to one for all $n_c > 1$, and it is not shown here. 
    \label{fig2}}
\end{figure*}

As before, we focus on the probability that the flock is in a single connected cluster, $\mathpzc{P}_{c}$; results are shown in Fig. \ref{fig2}.  In the topological case, this probability is one for all values of $n_c$, both in the case where graphs are chosen at random ($\mathpzc{P}_c^{\rm{ran}}$) and when the graphs are chosen from the motional distribution ($\mathpzc{P}_c^{\rm{mot}}$). In contrast, for the metric case the connection probability for random graphs  $\mathpzc{P}_c^{\rm{ran}}$ is close to one for only for $n_c \gtrsim 10$, while even larger values of $n_c \gtrsim  15$ are needed for the connection probability from the motional potential to be close to one. 

We note that our simplified MC method includes graphs which do not have a three-dimensional layout:  while any configuration $x$ can be mapped onto a graph $G$ (Fig. \ref{fig1}), not all graphs $G$ correspond to a configuration $x$. If we break the graph into disconnected pieces, it is always easier to find a mapping into a configuration $x$, and so we expect that sampling graphs will overestimate the probability that flocks are connected.  This is confirmed by a detailed comparison between  Figs. \ref{fig5} and \ref{fig2}.  Despite this quantitative difference, the MC on graphs confirms the qualitative scenario from the MC on positions: the metric potential has a strong repulsive effect which rips the flock apart into multiple components.  

\begin{figure}[b]
  \centerline{\includegraphics[width=0.975\linewidth]{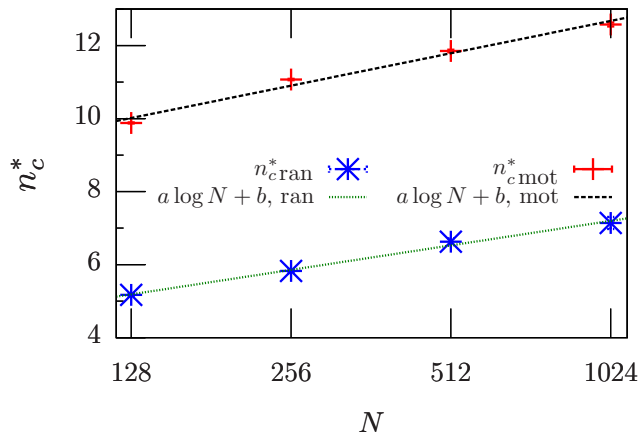}}
  \caption{
    Crossover values $n_c^\ast$ of the number of neighbors in the metric case, given by $\mathpzc{P}_c (n_c^\ast, N) = 1/2$, for random graphs and for graphs drawn from the motional potential, as functions of the flock size $N$. The fitting functions $n_c^\ast = a \log N + b$ are shown to guide the eye. 
    \label{fig3}}
\end{figure}

Figure \ref{fig5} gives a hint that, with metric interactions, larger flocks have a lower probability of being connected, and this trend continues to larger $N$ for the simulation on graphs in Fig. \ref{fig2}.  Put another way, larger flocks require a larger range of interaction $n_c$ in order to stay connected.  Marking the crossover $n_c^\ast$ between connected and disconnected regimes by $\mathpzc{P}_c^{\rm{mot}} (n_c^\ast ,N) = 1/2$, we see in Fig. \ref{fig3} that $n_c^\ast$ increases with $N$ along an approximately logarithmic trajectory, such that a factor of two increase in $N$ requires an extra contact to maintain coherence.  

\section{Discussion}\label{disc}

The past decade has seen considerable interest in the use of maximum entropy models to describe biological networks, from single protein molecules up to groups of organisms.  The maximum entropy method has deep connections to equilibrium statistical mechanics:  These connections are a source of intuition, but also create opportunities for confusion.   The derivation of entropic forces thus requires some care.

The maximum entropy--distribution that is consistent with pairwise correlations among the variables in a network (e.g. the normalized velocities of the birds) has the form of a Boltzmann distribution in which the ``energy'' is built out of pairwise interactions among these variables.    This is not, of course, the actual energy, and there is no reason to think that the interactions out of which the energy is built correspond to microscopic interactions.  If, as in the case of flocks, we can build a successful maximum entropy model by matching only local correlations \cite{bialek2012statistical,bialek2014speed,cavagna+al2014short-vs-long}, it is plausible that the underlying interactions are local, but the precise statement is that the full correlation structure in the flock as a whole is the minimal consequence of the local correlations.  

In the case of proteins, maximum entropy methods have been used to describe the ensemble of amino--acid sequences that are consistent with being a member of a particular protein family, and hence having a particular three--dimensional structure \cite{bialek2007rediscovering,seno2008,weigt2009identification,marks2011protein,sulkowska+al2012}.  This is the inverse of the usual protein folding problem, where we are given the sequence and asked to predict the structure.  In the (forward) folding problem, it is widely believed that interactions are local and that the structures we see often really are at thermal equilibrium.  Approximate solutions of the inverse problem show that the effective interactions in the maximum entropy model extend over a much shorter range than the correlations in amino--acid substitutions, to the point that one can identify physical contacts \cite{weigt2009identification} and hence go quite some way toward structure prediction from the sequence ensemble alone \cite{marks2011protein,sulkowska+al2012}.  

If we could write a successful maximum entropy model for the joint distribution of amino--acid sequences and the associated protein structures, then with the sequence held fixed it should reduce to a Boltzmann distribution over structures, with local interactions.  But if we sum over sequences to obtain the distribution of structures, then---by the arguments in Section \ref{entforces}---there will be an effective entropic potential that favors structures which can be stabilized by many different amino acid sequences.   These are precisely the ``highly designable structures'' identified long ago by Li and colleagues \cite{li+al_96,li+al_98}.

In a flock of birds, there is no part of the problem that is in thermal equilibrium, but nonetheless we can   write a maximum entropy approximation to the joint distribution of positions and velocities for all the birds in the flock, as in Eqs. (\ref{pxs1}) or (\ref{pxs2}).  Once we integrate out the velocities, the resulting motional distribution of positions has a term in the exponential that is exactly the logarithm of the partition function for the velocities at fixed positions---the free energy.  For flocks that are strongly polarized, as in real flocks, this free energy is dominated by the entropy of the birds' velocities, and in this sense the flock is subject to entropic forces.   We expect these entropic forces to be repulsive, since disconnected groups of birds have more freedom to reorient their flight directions, and this intuition is borne out by detailed simulations.  The surprise is that strength of this repulsion depends dramatically on the form of the correlations that we constrain.

If we imagine that the essential correlations are between a bird and its $n_c$ nearest neighbors (topological interactions), then the entropic forces are quite weak, and leave flocks fully connected with high probability at reasonable values of $n_c$.  In contrast, if the essential correlations are between a bird and all the other birds within  fixed distance $r_c$ (metric interactions), then the entropic forces are so large that flocks are almost always disconnected.   

Given the complicated form of the motional distribution, Monte Carlo (MC) simulations are computationally demanding, and they are thus limited to  small flocks. To address this problem, we explored a slightly different formulation in which we sample graphs of bird--bird interaction rather than the positions of the birds themselves; this allows  for using an efficient MC update algorithm based on LDL matrix factorization \cite{davis1999modifying}, with which we could analyze realistically large flocks.   Our main result is in line with the one obtained by sampling the positions: For a topological network, the configurations generated by the motional distribution are connected for all values of the number of nearest neighbors. On the other hand, for a metric network the positional configurations are disconnected with high probability unless the 
metric interaction range is increased to unrealistically large values. 

In real flocks, the absence of large local density fluctuations means that the distinction between topological and metric interactions must be based on comparisons across different flocking events \cite{bialek2012statistical}.  Our analysis of entropic forces provides a different path to comparing these two models.  While the strongly repulsive entropic effects of metric interactions could be compensated by explicit cohesive forces, the fact that positional correlations in flocks are weak \cite{cavagna2008gofr} means that these strong opposing forces would have to be carefully balanced.  Such fine tuning is not needed if the essential correlations are topological rather than metric.  

\begin{acknowledgments}
MC is grateful to TA Davis and WW Hager for advice and support on sparse LDL factorizations. Work at Princeton and CUNY was supported in part by NSF Grants  PHY--1305525 and CCF--0939370, by the Human Frontiers Science Program, and by the Simons and Swartz Foundations.  Work in Rome was supported in part by IIT grant Seed Artswarm, ERC--StG Grant No. 257126, and US--AFOSR Grant No. FA95501010250 (through the University of Maryland).
\end{acknowledgments}

\appendix 

\section{Monte Carlo update with LDL factorization}\label{app1}

Here we discuss the MC simulations with the  probability $\mathpzc{p}_{\rm{mot}}(G)$ from Eq. (\ref{por}).  In Appendix \ref{app1A} we show that $\mathpzc{p}_{\rm{mot}}(G)$ can be related to the LDL factorization of the Laplacian matrix (\ref{eq11}), and in Appendix \ref{app1B} we show that a MC step can be performed efficiently by using a known update algorithm for LDL factorizations.   For the sake of simplicity, we  consider the metric case, and we assume that $G$ is connected. The results below can then be  easily extended to the topological case and to graphs with multiple connected components. 

\subsection{Relation between the motional probability and the Laplacian LDL factorization}\label{app1A}

Since $G$ has only one connected component, there is a single zero eigenvalue of $\Lambda$ \cite{caughman2006kernels}, which we denote by $\lambda_1$, and Eq. (\ref{por}) shows that the probability $\mathpzc{p}_{\rm{mot}}(G)$ is determined by the product of the nonzero eigenvalues of the Laplacian (\ref{eq11}). Here, we will show that this product is related to the LDL factorization of $\Lambda$ \cite{davis2013user},  which reads
\be\label{eq10}
A \equiv P \Lambda P^T = L D L^T, 
\ee
where in  Eq. (\ref{eq10}) $P$ is a permutation matrix, $L$ is a lower-triangular matrix with unit diagonal elements, $D$ is a diagonal matrix whose diagonal elements will be denoted by $\{ d_i \}$, and the matrix $A$ has the same eigenvalues as $\Lambda$. Since $\Lambda$ has one zero eigenvalue, there is a single zero diagonal entry in $D$, which we denote by $d_1$.  We will now establish the connection between the spectrum of $\Lambda$ and its LDL factorization by proving the following identity
\be\label{eq12}
\prod_{i=2}^N \lambda_i = N \prod_{i=2}^N d_{i}.
\ee

To prove Eq. (\ref{eq12}), let us consider the characteristic polynomial of $A$: by using Sylvester's theorem, we have
\be\label{eq72}
f(\lambda)  \equiv  \det(A - \lambda I)  
 =  \det(DL^TL - \lambda I). 
\ee
The characteristic polynomial (\ref{eq72}) can be rewritten as 
\bea\label{eq40}
f(\lambda) = a_N - \lambda a_{N-1} + \cdots + (-1)^N \lambda^N,
\eea
where $a_i$ is the sum of all diagonal minors of $D L^TL$ containing $i$ rows and $i$ columns \cite{jacobson1985basic}. Since $d_1 = 0$, the first row in $D L^TL$ is zero, thus the only nonzero diagonal minor with one row and one column is the one obtained by deleting the first row and the first column from $D L^TL$. It follows that 
\be\label{eq33}
a_{N-1} = \det(B),
\ee
where $B$ is a $(N-1)\times (N-1)$ matrix with entries $B_{ij} = \sum_{l=1}^N D_{il}(L^TL)_{lj}$, $i,j = 2, \cdots, N$. Since $D$ is diagonal,  $B$ is given by the product of the matices obtained by removing the first row and the first column in $D$ and $L^TL$ respectively: hence, from Eq. (\ref{eq33}) we obtain
\be\label{eq41}
a_{N-1} = \left( \prod_{i=2}^N d_i \right) \det( C_{1 \, 1}),
\ee
where $C_{1 \, 1}$ denotes the matrix obtained from $L^TL$ by deleting the first row and the first column.

Since the eigenvalues of $A$ are $0 = \lambda_1 < \lambda_2 \leq \cdots \leq \lambda_N$,  the eigenvalues of $A-\lambda I$ are $- \lambda, \lambda_2 - \lambda, \cdots, \lambda_N -\lambda$. Thus, the characteristic polynomial (\ref{eq72}) reads 
\bea\label{eq70}
f(\lambda) & = & (-\lambda) (\lambda_2 - \lambda) \cdots (\lambda_N -\lambda)\\ \nn
& = & - \lambda\,  \prod_{i=2}^N\lambda_i + O(\lambda^2). 
\eea
Comparing the coefficient of $\lambda$ in the right--hand side of Eq. (\ref{eq40}) with that in the right--hand side of Eq. (\ref{eq70}) and using Eq. (\ref{eq41}), we obtain 
\be\label{eq43}
\prod_{i=2}^N\lambda_i = \det( C_{1 \, 1}) \prod_{i=2}^N d_i . 
\ee

To derive Eq. (\ref{eq12}), let us compute $\det( C_{1 \, 1} )$. Equations (\ref{eq11}), (\ref{eq10}) show that the vector  $u \equiv (1,\cdots, 1)$ is an eigenvector of $A$ with eigenvalue zero. Setting $e_1 \equiv (1,0, \cdots, 0)$, we have that $(L^T)^{-1} e_1$ is also an eigenvector of $A$ with eigenvalue zero:
\bea\label{eq30}
A (L^T)^{-1} e_1 & = & L D L^T (L^T)^{-1} e_1 \\  \nn
& = & L D e_1\\  \nn
& = & 0,
\eea
where in the second line of Eq. (\ref{eq30}) we have $D e_1 = 0 $ because $d_1 = 0$.  Given that $\Lambda$ is symmetric, the geometric multiplicity of $\lambda_1$ is equal to its algebraic multiplicity, the latter being equal to one. It follows that there is only one eigenvector of $\Lambda$ with zero eigenvalue, thus only one eigenvector of $A$ with zero eigenvalue: hence, $u$ must be proportional to $(L^T)^{-1}e_1$. Also, $(L^T)^{-1}$ is upper triangular with unit diagonal entries, thus the first component of $(L^T)^{-1}e_1$ is equal to one, implying that
\be\label{eqeu}
u = (L^T)^{-1}e_1. 
\ee
To prove Eq. (\ref{eq12}), we will  relate the norm of $u$ to $L$:
\bea\label{eq32}
N & = & u^T u  \\ \nn
& = & e_1^T L^{-1}  (L^T)^{-1} e_1 \\ \nn
& = & e_1^T (L^TL)^{-1} e_1 \\ \nn
& = & [(L^TL)^{-1}]_{1\,1},
\eea
where in the first line of Eq. (\ref{eq32}) we used Eq. (\ref{eqeu}), and $[(L^TL)^{-1}]_{1\,1}$ denotes the entry in the first row and first column of $(L^TL)^{-1}$. By using Cramer's rule, the last line in Eq. (\ref{eq32}) can be rewritten as
\be\label{eq31}
[(L^TL)^{-1}]_{1\,1}  = \frac{\det(C_{1\,1})}{\det(L^TL)} 
 =  \det(C_{1\,1}), 
\ee
where  we use the identity $\det(L^TL) = [\det(L)]^2 = 1$. Equations (\ref{eq32}) and (\ref{eq31}) imply that
$\det(C_{1 \, 1}) = N$.
Substituting into Eq. (\ref{eq43}),  we obtain Eq. (\ref{eq12}). 
\vfill

\subsection{Monte Carlo with LDL-factorization update}\label{app1B}

Since we intend to sample  the space of graphs with a  constant total number of edges, a MC move is given by one edge insertions and one edge deletion. We use Eqs. (\ref{eq11}) and (\ref{eq13}) to rewrite the Laplacian of $G$ as  
\be\label{eq14}
\Lambda_{ij} = \frac{J}{n_c} \left[ \left( \sum_{l=1}^N n_{il} \right)\delta_{ij} - n_{ij}\right]. 
\ee
We then take two vertices $i$ and $j$ in $G$ that are not connected, and we insert an edge between them. As a result, we obtain a new graph $G'$ with Laplacian $\Lambda'$,
\be\label{eq17}
\Lambda' = \Lambda + \frac{J}{n_c}\,  v \cdot v^{T},
\ee
where the vector $v$ is given by $v_l = \delta_{il} - \delta_{jl}$. Since $\Lambda$ is sparse and $\Lambda'$ is related to $\Lambda$ by a transformation of the form (\ref{eq17}),  it can be shown  \cite{davis1999modifying} that the LDL factorization of the Laplacian $\Lambda' = L' D' {L'}^{T}$ can be computed from the LDL factorization  of $\Lambda$ in a number of steps 
proportional to the number of nonzero entries in $L$ that change upon the update, which is bounded above by $O(N^2)$. The probability $\mathpzc{p}_{\rm{mot}}(G')$ is then obtained from $D'$ according to Eq. (\ref{eq12}). An edge insertion can be thus performed with not more than $O(N^2)$ operations; by the same argument, an edge deletion---and thus a full MC move---can be also performed with this number of operations.

Finally, it is important to point out that  replacing the denominators $n_i, n_j$ in Eq. (\ref{eq55}) with their average value $n_c$---see Section \ref{results4.3}---is crucial for this efficient update method to work: indeed, without this simplification the Laplacian would not have the simple form (\ref{eq14}), and a Laplacian update upon edge insertion or deletion would not be of the form (\ref{eq17}), thus preventing  us from using the LDL factorization update algorithm above.

\bibliographystyle{unsrt}
\bibliography{bibliography}

\begin{thebibliography}{10}
\bibitem{note1}
In Eq. (\ref {Psgivenx}), with topological interactions we have $n_i(x) = n_c$
  for all $i$, and hence \begin {equation} P(s|x) = {1\over {Z(x;J)}} \exp
  \left [ {J\over {n_c}}\sum _{i=1}^N \sum _{j=1}^N n_{ij}(x) {\vec s}_i \cdot
  {\vec s}_j \right ] . \end {equation} Comparing with Ref. \cite
  {bialek2012statistical}, we see that what was $J$ in the previous work is
  $J/n_c$ in the current formulation. In the metric case, we compare Eq. (\ref
  {Psgivenx}) with the flock configurations in Ref. \cite
  {bialek2012statistical} with a nearly homogeneous density $\rho $ by setting
  $n_i(x) \approx n_c = 4\pi \rho \, r_c^3 / 3$, and we obtain the same
  relation between the interaction strengths.
\end{thebibliography}
\end{document}